\documentclass[a4paper,fleqn]{manuscript/cas-dc}

\usepackage[numbers]{natbib}
\usepackage{xcolor}
\usepackage{amsmath}
\usepackage[FIGTOPCAP]{subfigure}
\usepackage{graphicx}
\usepackage{dcolumn}
\usepackage{bm}

\def\tsc#1{\csdef{#1}{\textsc{\lowercase{#1}}\xspace}}
\tsc{WGM}
\tsc{QE}
\tsc{EP}
\tsc{PMS}
\tsc{BEC}
\tsc{DE}

\begin{document}
\let\WriteBookmarks\relax
\def\floatpagepagefraction{1}
\def\textpagefraction{.001}
\shorttitle{}
\shortauthors{.}

\sloppy 

\title [mode = title]{Predicting Mechanical Properties from Microstructure Images  in Fiber-reinforced Polymers using Convolutional Neural Networks}                      

\author[1]{Yixuan Sun}[orcid=0000-0003-1109-3380]
\credit{Methodology, Software, Validation, Formal analysis, Data Curation, Visualization, Writing - Original Draft}

\author[2]{Imad Hanhan}[orcid=0000-0003-3931-5117]
\credit{Conceptualization, Methodology, Investigation, Data Curation, Visualization, Writing - Review \& Editing}

\author[2]{Michael D. Sangid}[orcid=0000-0002-1986-8673]
\credit{Conceptualization, Investigation, Resources, Writing - Review \& Editing, Supervision, Project administration, Funding acquisition}

\author[3]{Guang Lin}[orcid=0000-0002-0976-1987]
\cormark[1]
\ead{guanglin@purdue.edu}
\credit{Conceptualization, Investigation, Methodology, Resources, Writing - Review \& Editing, Supervision, Project administration, Funding acquisition}

\address[1]{School of Mechanical Engineering, Purdue University, West Lafayette, IN, 47907, USA}
\address[2]{School of Aeronautics and Astronautics, Purdue University, West Lafayette, IN, 47907, USA}
\address[3]{Department of Mathematics, School of Mechanical Engineering, Department of Statistics (Courtesy), Department of Earth, Atmospheric, and Planetary Sciences (Courtesy), Purdue University, West Lafayette, IN 47907, USA}

\cortext[cor1]{Corresponding author, Fax: 765 494 0548; Tel: 765 494 1965}

\begin{abstract}
Evaluating the mechanical response of fiber-reinforced composites can be extremely time consuming and expensive. Machine learning (ML) techniques offer a means for faster predictions via models trained on existing input-output pairs and have exhibited success in composite research. This paper explores a fully convolutional neural network modified from StressNet, which was originally for linear elastic materials, and extended here for a non-linear finite element (FE) simulation to predict the stress field in 2D slices of segmented tomography images of a fiber-reinforced polymer specimen. The network was trained and evaluated on data generated from the FE simulations of the exact microstructure. The testing results show that the trained network accurately captures the characteristics of the stress distribution, especially on fibers, solely from the segmented microstructure images. The trained model can make predictions within seconds in a single forward pass on an ordinary laptop, given the input microstructure, compared to 92.5 hours to run the full FE simulation on a high-performance computing cluster. These results show promise in using ML techniques to conduct fast structural analysis for fiber-reinforced composites and suggest a corollary that the trained model can be used to identify the location of potential damage sites in fiber-reinforced polymers.
\end{abstract}


\maketitle

\section{Introduction}
Fiber-reinforced composites offer a transformative opportunity for the manufacturing of lightweight structures.  One imposing difficulty to the widespread adoption of these fiber-reinforced composites is the substantial time and cost necessary for certifying their structural integrity in engineering applications. Specifically, uncertainties regarding how the complex microstructure and defects within these materials influence the mechanical response have impeded their usage in structural applications.  As discussed in the Materials Genome Initiative, a primary goal is to reduce the time and cost to bring new materials and structures to market by 50\% \cite{national2011materials}.  To enable this goal, the NASA 2040 Vision discusses opportunities and needs for tool maturation, specifically articulating the importance of the development, verification, and validation of multiscale modeling approaches \cite{liu2018vision}.  Towards this objective, this paper identifies rapid and efficient methods to predict the mechanical response of fiber-reinforced composite materials with emphasis on their microstructure, using machine learning techniques.  

In composite materials, one approach that can offer reduced computational costs for microstructure informed modeling is typically based on homogenization schemes or unit cell approaches.  Broadly speaking, homogenization schemes rely on simulating the micromechanical behavior of a representative volume of the material and use the resulting material response to inform hierarchical modeling schemes at coarser length-scales \cite{kammoun2011first, aboudi1995micromechanical, poulios2016homogenization, fliegener20143d}.  In a similar fashion, the method of cells provides a periodically repeating unit of the material’s microstructure and applies the continuity of the tractions and displacements over these cells to construct a continuum response of the solid \cite{aboudi1989micromechanical, aboudi2013generalized, bohm2002multi, modniks2013modeling}.   In such modeling schemes, a reduction in the dimensionality of the modeling results in a loss of information, and as a consequence, these approaches are not able to capture local and rare events of damage. Further, these schemes can be even more challenging for a thermoplastic matrix, since the matrix experiences significant non-linear deformation which can be expensive to simulate. Therefore, these modeling events may not be appropriate to capture incipient damage.  

Based on advancements in high-resolution characterization techniques coupled with microstructural sensitive modeling, a physics-based understanding and eventual prediction of the events that result in local damage is possible \cite{sangid2019coupling}; two examples are given here, which focus on damage initiating at an individual fiber, such as incipient damage or kink formation.  For damage of individual fibers in continuous fiber-reinforced composites, tomography has been used to study reorientation, breakage, and kink formation \cite{emerson2017individual, garcea2017mapping, emerson2018quantifying, wang2017x}, which has led to the development of models at the fiber level \cite{wind2014comparison, davidson2020micromechanics}.  Additionally, the formation and growth of incipient damage have been detected at the fiber tips of short discontinuous fibers, as the primary failure mechanism \cite{sato1991microfailure, rolland2016damage, hu2016real, hanhan2020predicting}.  Through detailed modeling, Hanhan et al.~individually meshed the microstructural features and determined that high hydrostatic stress was responsible as the primary driver for damage.  In their analysis, the predictions for the site of damage through finite element (FE) simulations were in agreement with the same sites observed via \textit{in-situ} experimental observations \cite{hanhan2020predicting}.  While segmenting the fiber and void features identified through tomography and explicitly meshing these features for use within a FE model has provided promising results, it required 44.5 million elements to model a volume of ~0.0096 $mm^3$ \cite{hanhan2020predicting}, which makes this method computationally impractical within an engineering workflow.   For this reason, machine learning offers a promising direction for the exploration of computationally efficient tools and approaches. 

Due to their extraordinary ability to learn non-linear mapping and their advances in computer vision tasks, deep neural networks, and a class of deep neural networks, convolutional neural networks (CNNs), have shown promising results in engineering research. CNNs have been widely used in solving engineering problems since AlexNet first used a deep convolutional network to perform ImageNet classification tasks \cite{krizhevsky2012imagenet}. Characteristics of CNNs, such as translation invariant and weight sharing, allow them to perform well for computer vision tasks. Fully connected neural networks were trained and evaluated on pairs between selected material treatment conditions or design variables and outputs to predict the compressive strength of heat-treated woods and concrete \cite{tiryaki2014artificial, khademi2017multiple, young2019can}. However, since the information of composite materials can not be naturally stored in vectors that are mutually independent, careful feature selection is imperative to training a successful fully connected neural network. In recent studies, to relax this requirement, because of their parameter-sharing and translation invariant characteristics, CNNs have been utilized for predicting material properties by looking directly at the material topology images. Gu et al.~built a CNN to predict the toughness of composites based on the topology of base materials \cite{gu2018novo}. Yang et al.~adopted a 3D CNN to predict the stiffness of high contrast composites from generated microscale volume elements \cite{yang2018deep}.  Hanakata et al.~used a traditional CNN structure to estimate the yield stress and strain based on graphene containing their kirigami designs \cite{hanakata2018accelerated}. While these studies showed high accuracy in the evaluation of the CNN models, they mainly focused on predicting a single value from the microstructure data. 

In order to investigate incipient damage and kink formation, it is of interest to utilizing deep learning methods to quickly obtain the stress field of a composite. Khadilkar et al.~developed a two-stream CNN to predict the stress field for a stereolithography 3D printing process, in which the network outputs a long vector of $O(10^{5})$ dimension that was reshaped to match the size of 2D slices of the specimen \cite{khadilkar2019deep}. Nie et al.~developed a fully convolutional network, StressNet, with an encoder-decoder structure to predict the stress field of linear elastic cantilevered structures in an end-to-end manner with additional injected load information and displacement boundary conditions \cite{nie2020stress}. This network is able to preserve the structural information of the input while extracting higher representations for accurate prediction. 

In predicting the stress field of material given the microstructure, CNN-based deep learning models are suitable candidates because of their ability to capture hierarchical or structural information embedded in the microstructure. Based on its success in the linear elastic cantilevered structures, in this work, we adopted and simplified the structure of StressNet, and extended it for a non-linear FE simulation to investigate its generalization ability on fiber-reinforced polymers. We only took the 2D segmented microstructure images as the network input without additional information such as load or boundary conditions. More specifically, the convolutional layers were regarded as feature extractors that slide through the input and do the discrete convolution operations, generating feature maps containing high-level representations. During this process, the same convolutional filter was used for the entire input so that the trainable weights were shared throughout the input. This limited the number of trainable parameters that can make the training process more efficient. The downsampling operations in the pooling layers achieved translational invariance, making the absolute location of a certain pattern less important.  A dataset containing 5321 2D microstructure slices sampled from segmented X-ray tomography images of a composite specimen and its corresponding FE simulation were used to train and validate the network. The trained model can make local stress predictions in a single forward pass from the given microstructure images within seconds on a laptop, compared to 92.5 hours to run the full FE simulation on a high performance computing cluster. According to the evaluation results, the trained network is able to output the stress distribution and capture important characteristics, especially on fibers, from the corresponding segmented microstructure 2D slices. The rest of the paper is organized as follow: Sections \ref{exp} and \ref{fem} describe the data generation process; Section \ref{ml} covers the workflow for training and evaluating the neural network; Section \ref{results} shows the evaluation results and discussion; Section \ref{conclusion} summarizes this work.

\section{Experiment}
\label{exp}
\begin{figure}
    \centering
    \includegraphics[width=0.9\linewidth]{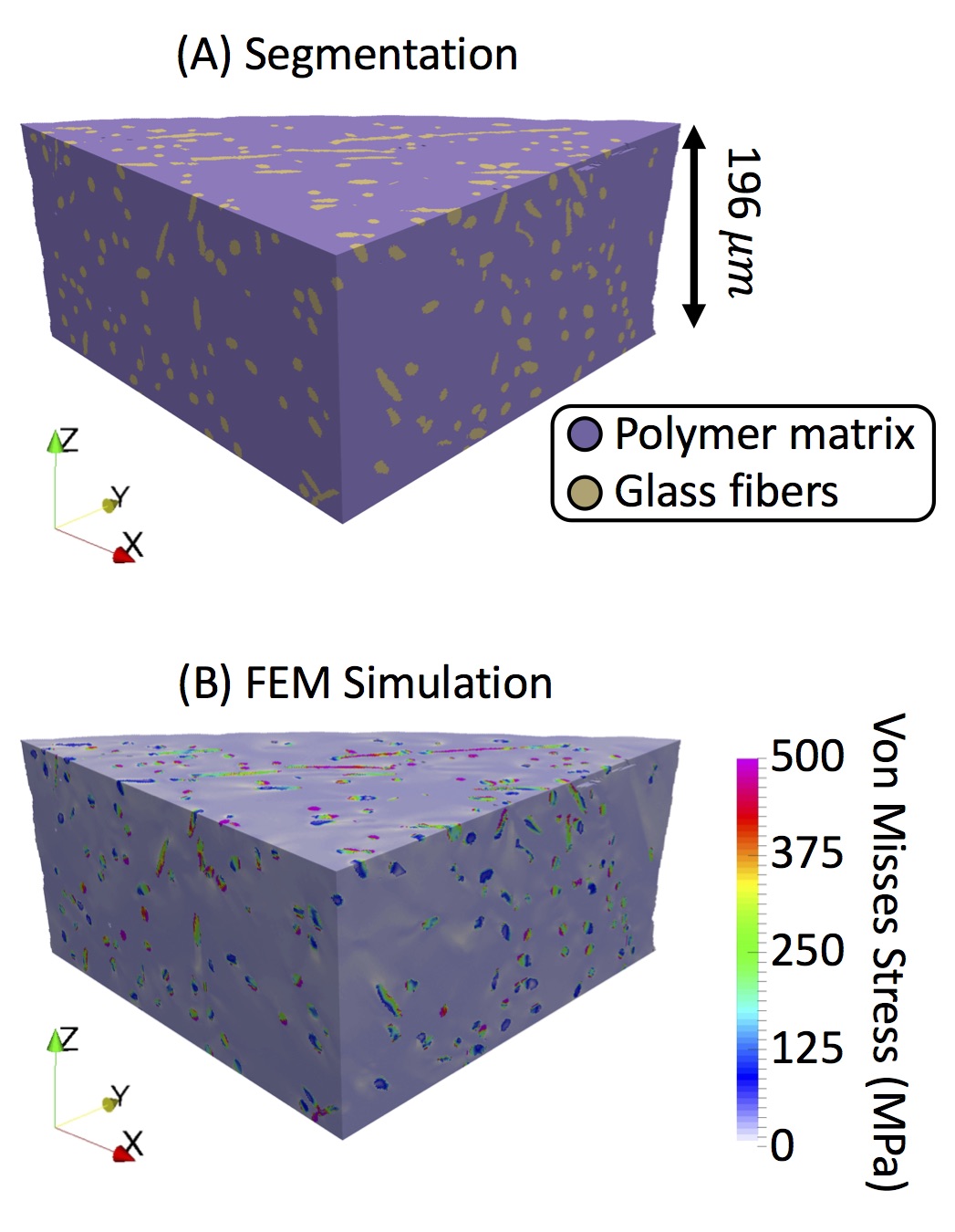}
    \caption{The segmented specimen microstructure is shown in (A), where the blue state represents polymer and the yellow state represents fibers, and the FE simulation result is shown in (B), where the Von Mises stress is plotted.}
    \label{seg}
\end{figure}

In order to provide a dataset for sampling that can be used for both neural network training and testing, the exact microstructure of a fiber-reinforced thermoplastic composite was analyzed. Specifically, the material studied through FE modeling was an injection-molded composite where the polymer matrix was polypropylene, and the fiber fillers were E-glass fibers. The E-glass fibers were approximately 10 $\mu m$ in diameter, and had varying lengths and orientations in the final composite specimen, due to the injection molding process~\cite{bailey1987study}. The injection molded part was a cylindrical rod with a diameter of 1.27 cm and a length of 45.72 cm, where the injection molding direction was in the length direction of the cylinder (Z-axis in Figure~\ref{seg}). The cylindrical rod was then machined into a smaller dog-bone shaped specimen with a gauge section diameter of 2.5 mm, and a gauge section length of 5 mm.

In order to extract the 3D properties of the microstructure, as well as create a 3D FE model that can compute the local stresses within the microstructure, X-ray micro-computed tomography was conducted. The data acquisition was conducted at the Advanced Photon Source in Argonne National Laboratory at beam-line 2-BM using synchrotron X-rays. X-ray projections were acquired using X-ray energy of 25 keV and a detector exposure time of 100 ms. The detector used for capturing the X-ray projections was placed 75 mm downstream from the specimen. The specimen was rotated at 0.5$^o/s$ through a range of 180$^o$, where an X-ray projection was acquired every 0.12$^o$. The total 1500 X-ray projections were reconstructed using TomoPy~\cite{Gursoy2014} into a 3D image volume with dimensions 3.33 by 3.33 by 1.61 $mm$ (with a pixel size of 1.3 $\mu m$).

\section{Finite Element Simulation}
\label{fem}

The 3D image volume required several image processing steps in order to extract the microstructural features of interest for the FE simulation. Specifically, a total of four feature types needed to be detected and extracted: the exterior edge of the specimen, the glass fibers, the porosity, and the polymer matrix. The exterior edge of the specimen was detected using an in-house MATLAB algorithm which uses an initial guess of the center and radius of the specimen. This is followed by automated image processing which maps the image intensity values from a range of [0,1] to a range of [0.4, 0.9], converts the image into a binary image using a threshold of 0.655 of the median intensity of the image, dilates the binarized image using a disk structural element with a radius of 4 pixels, and finally adjusts the binarized image by filling any holes~\cite{hanhan2020predicting}. The specimen edge detection was verified using ModLayer~\cite{Hanhan2019b}.

The glass fibers were detected using an iterative and supervised 2D and 3D combined algorithm~\cite{Agyei2018a}. Next, the porosity was detected using a combination of Weka machine learning segmentation~\cite{Arganda-Carreras2017} and manual correction using ModLayer~\cite{Hanhan2019b}. Lastly, voxels within the interior of the specimen which were not classified as fibers or pores were labeled as the polymer matrix. The final detection of all the microstructural features was verified using ModLayer~\cite{Hanhan2019b}. A region of interest was virtually down-selected and can be seen in Figure~\ref{seg}A. This region was chosen because it was determined to be a critical region that exhibited significant experimental damage within the microstructure~\cite{hanhan2020predicting}.

The meshing of this down-selected region was conducted in ParaView. First, voxels that were characterized as porosity were removed from the 3D volume, effectively creating gaps within the polymer matrix. The remaining voxels (which corresponded to the fibers and the matrix) were meshed using tetrahedral elements directly from the voxelated microstructure, which resulted in ideal and geometrically identical tetrahedral elements~\cite{hanhan2020predicting}. In total, the down-selected region of interest shown in Figure~\ref{seg} contained 44.5 million elements.

The meshed fiber elements were assigned linear elastic mechanical properties with an elastic modulus of 72.4 GPa and a Poisson's ratio of 0.2~\cite{agarwal1990,bauccio1994asm}. The meshed matrix elements were assigned non-linear mechanical behavior properties using a multi-linear isotropic hardening model~\cite{Mohammadpour2014}. The positive X, negative Y, and negative Z surfaces (Figure~\ref{seg}) were assigned roller boundary conditions. The positive Z surface was displaced positively by 7.8 $\mu m$~\cite{hanhan2020predicting}. The free surface of the specimen was allowed to remain a free surface with no boundary condition. The FE simulation was solved in Abaqus using 300 parallel processors with 1.92 TB of memory in 92.5 hours. A sample of the result of the simulation is shown in Figure~\ref{seg}B, where Von Misses stress is plotted. The result of the simulation has been validated by Hanhan et al.~where they showed that locations of high stress spatially matched with locations of local experimental damage initiation (in the form of micro-void nucleation)~\cite{hanhan2020predicting}. For the machine learning application in this work, the segmented microstructure and the FE stress in the loading direction, $\sigma_{zz}$, were used for data sampling, training, and testing.

\section{Methodology: Deep Learning Framework}
\label{ml}
\subsection{Data Sampling}


Since the segmented microstructure shown in Figure~\ref{seg}A was represented by unit-less square voxels (with a voxel size of 1.3 by 1.3 by 1.3 $\mu m$), and the FE simulation was represented by tetrahedral mesh elements measured in $\mu m$, the stress field data was resized using the \texttt{scipy.ndimage.zoom} method via order 3 spline interpolation to match the segmented data. Data points were selected from the segmented microstructure and the corresponding stress field, with a sampling window size of $32 \times 32$ pixels. This particular sampling window size was chosen in order to (i) contain, on average, the entire fiber’s cross-section, (ii) generate as many samples as possible, and (iii) serve the downsampling purpose in the network training process. A sampled data point was stored in a rank 3 array of shape (32, 32, 2) with the first two axes indicating the spatial location of a voxel within the data point and the third axis representing the voxel microstructure type and stress value. During the sampling process, there was no overlapping between the sampling windows, which ensured that each data point was unique in the dataset. To investigate the importance of the sampling plane to the model performance, data points of equal size were sampled from three xy-plane, xz-plane, and yz-plane, respectively, and the data from different planes were used to train and evaluate the networks. 

\subsection{Preprocessing}

In order to evaluate the model performance on unseen data, the sampled data points were randomly split into a training set and a testing set with no overlapping in between. The training set contained 80\% of the entire dataset and the testing set contained 20\%. In addition, since the standardization or normalization of the input data can accelerate the training process, the input of the model was standardized. In this work, all the microstructure images from sampled data points were standardized using the sample mean and standard deviation with the same shape as the input in the training set. Specifically, the input data, $X$, for both the training set and the testing set was standardized as shown in Eq.~\ref{stand},

\begin{equation}
    X_{norm} = \frac{X - \mu_{train}}{\sigma_{train}}
\label{stand}
\end{equation}
where $\mu_{train}$ and $\sigma_{train}$ are the sample mean and standard deviation from the data points in the training set. It is important to note that if the trained model is used for new predictions, the new input data would need to be standardized in the same manner.

\subsection{Network Architecture}

\begin{figure*}
    \centering
    \includegraphics[width=\linewidth]{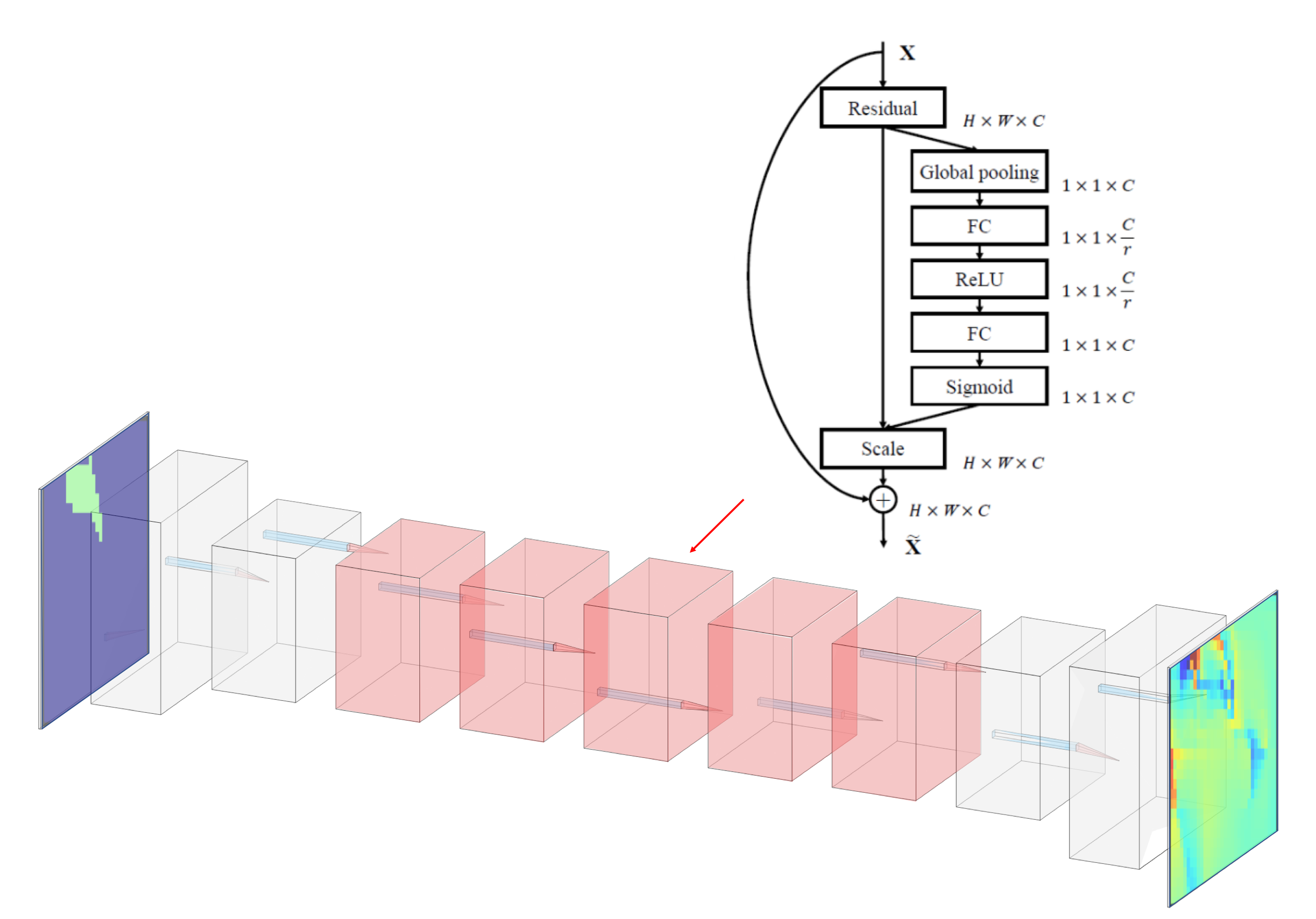}
    \caption{Convolutional neural network architecture with an encoder-decoder structure. This network takes microstructure images of size 32 $\times$ 32 as input and outputs the corresponding stress field of the same size. The red highlighted blocks are Squeeze-Excitation Residual blocks \cite{hu2018squeeze} and the rest are plain 2D convolution layers with MaxPooling.}
    \label{network}
\end{figure*}

A CNN-based network, shown in Figure \ref{network}, was adopted from StressNet \cite{nie2020stress} for predicting the stress field given the microstructure. StressNet was originally developed to identify the stress field of linear elastic materials. In this study, StressNet is adopted and modified to extend to the non-linear mechanical behavior of fiber-reinforced polymers. The network was comprised of an input layer, an output layer, and 11 hidden layers in between. 
The input was the segmented microstructure array of size $32\times32$, denoted by $X$. A typical convolutional block consists of one or more convolutional layers and a pooling layer. In this section, we describe the convolution operations and pooling operations, which are accompanied by the details of the network architecture in Table \ref{netart}.

\begin{table*}
\caption{Network Architecture Description} 
\centering 
\begin{tabular}{c c c c c c c} 
\hline\hline  
\multicolumn{2}{c}{\textbf{Operation Layers}}& \textbf{Number of Filters} & \textbf{Kernel Size} & \textbf{Stride} & \textbf{Padding} & \textbf{Output Size}\\ 
\hline  
\multicolumn{2}{c}{\textbf{Input Segmented Microstructure}} & - & - & - & - & $32 \times 32 \times 1$\\  
\hline  
\textbf{Convolution Layer} & ReLU & 32 & $3 \times 3$ & $1\times1$ & SAME & $32\times32\times32$ \\ 
\hline
\textbf{Pooling} & Max pooling & - & $2\times2$ & $2\times2$ & SAME & $16\times16\times16$ \\
\hline  
\textbf{Convolution Layer} & ReLU & 64 & $3 \times 3$ & $1\times1$ & SAME & $16\times16\times64$ \\ 
\hline
\textbf{Pooling} & Max pooling & - & $2\times2$ & $2\times2$ & SAME & $8\times8\times64$\\
\hline
\textbf{SE ResNet Layer} & ReLU  & 64 & $3\times3$& $1\times1$ &SAME &  $8\times8\times64$\\
\hline
\textbf{SE ResNet Layer} & ReLU & 64 & $3\times3$& $1\times1$ &SAME &  $8\times8\times64$\\
\hline
\textbf{SE ResNet Layer} & ReLU & 64 & $3\times3$& $1\times1$ &SAME &  $8\times8\times64$\\
\hline
\textbf{SE ResNet Layer} & ReLU & 64 & $3\times3$& $1\times1$ & SAME &  $8\times8\times64$\\
\hline
\textbf{SE ResNet Layer} & ReLU  & 64 & $3\times3$& $1\times1$ &SAME &  $8\times8\times64$\\
\hline
\textbf{Transposed Convolution} & ReLU & 64 & $3 \times 3$ & $2\times2$ & SAME & $16\times16\times64$ \\ 
\hline
\textbf{Transposed Convolution} & ReLU & 32 & $3 \times 3$ & $2\times2$ & SAME & $32\times32\times32$ \\ 
\hline
\textbf{Convolution Layer} & - & 1 & $3 \times 3$ & $1\times1$ & SAME & $32\times32\times1$ \\ 
\hline
\multicolumn{2}{c}{\textbf{Output Stress Field}} & - & - & - & - & $32 \times 32 \times 1$\\
\hline
\end{tabular}
\label{netart}
\end{table*}

 Feature maps are obtained after convolving the input array with filters. It can be expressed as follow:
\begin{equation}
S_i = W_i \star X  + b_i
\end{equation}
where $i$ denotes the filter number in a layer,  $S_i$ is the $i^{th}$ feature map generated, $W_i$ is the weight matrix associated with the $i^{th}$ filter, $b_i$ is the bias matrix for the $i^{th}$ filter, and $\star$ is the convolution operator. At the end of each convolutional layer, an activation function was applied on the feature maps. Activation functions introduce non-linearity into the network, enabling the approximation of non-linear underlying functions. A rectified linear unit (ReLU) was placed as the activation function for each feature map for its simplicity in calculating gradients and its ability to prevent gradient vanishing problems \cite{xu2015empirical}. The formulation of the ReLU was
\begin{equation}
ReLU(s) = \begin{cases}
s, & \text{if  } s > 0\\
0, & \text{Otherwise}
\end{cases}
\end{equation}
where $s$ was the entry of the feature map matrix $S_i$. The corresponding derivative with respect to $s$ was 
\begin{equation}
\frac{d ReLU}{ds} =
\begin{cases}
1, & \text{if  } s >0\\
0, &\text{Otherwise}
\end{cases}
\end{equation}
The final output of each convolutional layer was 
\begin{equation}
O = ReLU(S_i) = ReLU(W_i\star X + b_i)
\end{equation}

Pooling layers in a convolutional block was performed as a means to down-sample the feature maps. Depending on the pooling kernel size, stride, and padding options, the original feature maps were transformed into an array with a smaller size according to a certain criterion, such as maximum, minimum, average, or global average. The pooling operations were all max-pooling in our network, in which the maximum value in each pooling window was preserved, except the layers involving ResNet and SE blocks.

Among the hidden layers, some layers were made of residual blocks \cite{he2016deep} and squeeze-excitation blocks \cite{hu2018squeeze}. These layers increased the network's representation power by capturing identical mapping and leveraging the importance of different channels. 

\subsection{Network Training}
The implementation of the network was done in \texttt{Tensorflow v1.15}, a differential programming deep learning framework where the gradient-based optimization techniques, such as stochastic gradient descent (SGD), and Adam, can be easily achieved through auto-differentiation and back-propagation. The loss function was chosen as the mean squared loss (MSE) due to the nature of the regression problem. The optimizer for the network was Adam \cite{kingma2014adam}, which is a gradient-based, adaptive optimization method. Adam generally performs better than plain SGD for its usage of momentum leading faster convergence and ability to adaptively select a separate learning rate for each parameter as the training goes. A Tesla P100 GPU was utilized to train the network.

\subsection{Evaluation Metrics}
Due to the regression nature of the problem, we choose the coefficient of determination ($R^2$) as the metric for the model performance of the testing data. It is defined as 
\begin{equation}
R^2  = 1 - \frac{SS_{res}}{SS_{tot}}
\end{equation}
where $SS_{res} = \sum (f - y)^2$ and $SS_{tot} = \sum(y - \bar{y})^2$. Here, $f$ represents the predicted values from the CNN model, $y$ represents the true values in the FE dataset, and $\bar{y}$ is the sample mean of $y$. Since $R^2$ measures the fraction of the variance in the data that can be explained by the model, a perfect model is expected to have an $R^2$ of 1.

\section{Results and Discussion}
\label{results}

In order to create a model that effectively uses the segmented microstructure as an input, and generates $\sigma_{zz}$ as the output, models with the same network architecture were trained on sampled data points from different planes. Data points sampled from the xy-plane, yz-plane, and xz-plane were used to train the models, and the models were evaluated on the designated testing set. The sampled datasets from the aforementioned three orthogonal planes were of the same size. They contained 5321 data points each, in which 20\% were reserved for testing. All the networks were trained with 5000 epochs (the number of times that the network goes through the entire training set) with batch-normalization \cite{ioffe2015batch}. 

\begin{figure*}
\centering
\includegraphics[width=0.95\linewidth]{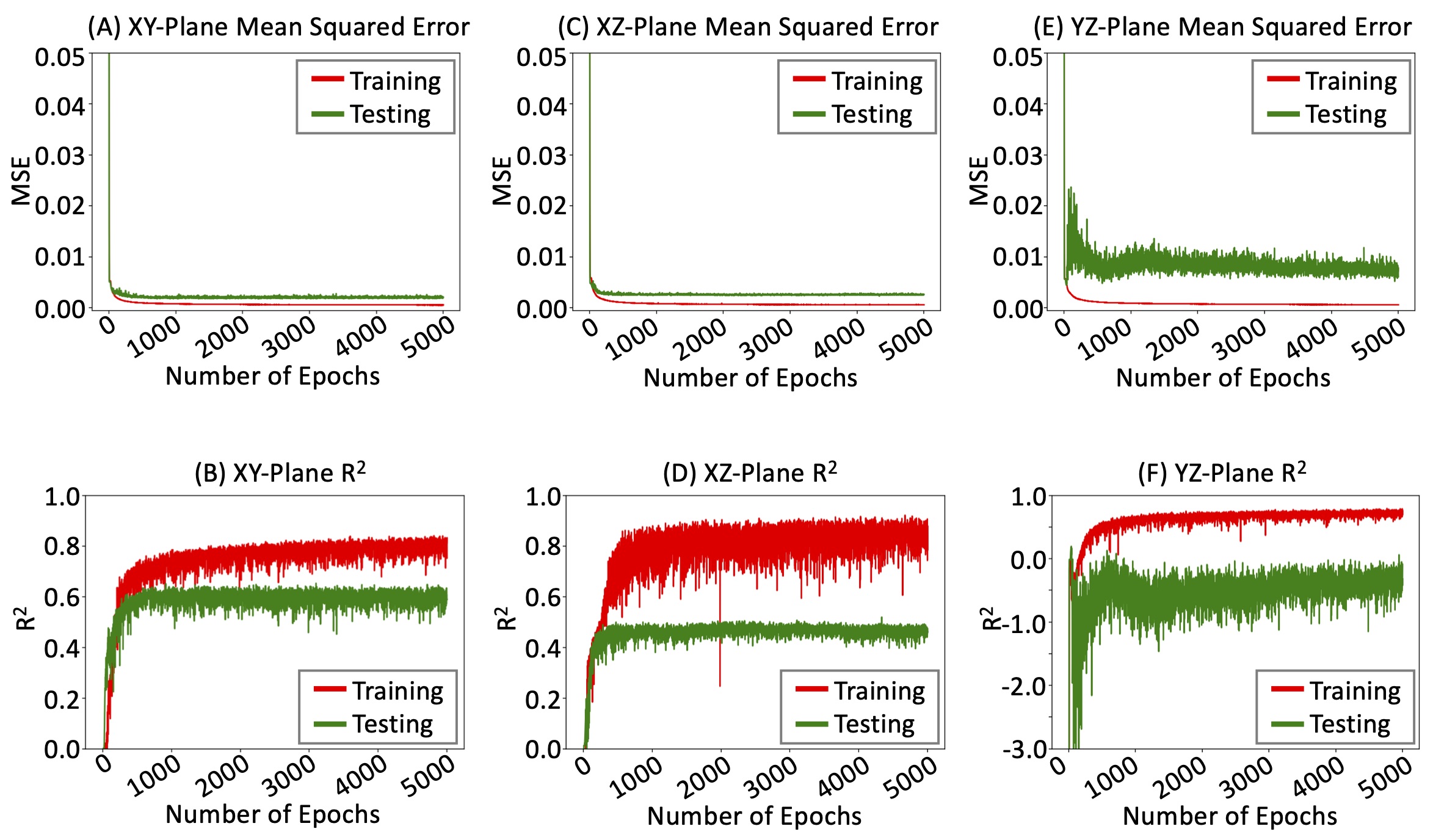}
\caption{Mean squared error of training and testing losses (top) and curves of coefficient of determination of model prediction on the training and testing sets (bottom) for each of the three orthogonal planes: (A,B) xy-plane, (C,D) xz-plane, and (E,F) yz-plane.}
\label{fig:MSE_and_R2}
\end{figure*}

Figures \ref{fig:MSE_and_R2}A and B show the model training process on the 2D slices sampled from the xy-plane. As the number of epochs increased, the MSE values from both training and testing decreased, and became stable after around 400 epochs, with a final testing error that was slightly higher than the training error. The opposite trend can be observed in the training process measured in $R^2$, where the scores on training and testing sets increased with the number of epochs. The curves also became flat after around 400 epochs, corresponding to a higher $R^2$ score for the training set compared to the testing set.

Figures \ref{fig:MSE_and_R2}C and D show the model training curves on the 2D slices sampled from the xz-plane. Similarly, the MSE values decreased as the number of epochs increased, while the gap between the training error and testing error was larger (after reaching stability) compared to the training process on the xy-plane data. In terms of the $R^2$ scores, although the score on the training set was comparable to that for the xy-plane shown in Figure \ref{fig:MSE_and_R2}B, the score on the testing set was much lower, making the data sampled from the xz-plane less reliable in predicting $\sigma_{zz}$.


  

The curves of the training process on slices sampled from the yz-plane are shown in Figures \ref{fig:MSE_and_R2}E and F. They generally follow the trend presented in the models trained on slices from the xy-plane and xz-plane.  However, the model performance on the testing set was evidently worse than the previous models, although the training loss and $R^2$ score on the training set were close.


  

\begin{table}
	\caption{The coefficient of determination ($R^2$) of model performance of predicting z-stress field on data sampled from different planes.}
	\centering
	\begin{tabular}{|c|c|c|c|}
		\hline
		Stages & xy-plane & xz-plane & yz-plane\\
		\hline
		Training & 0.88 & 0.82 & 0.80 \\
		\hline
		Testing & 0.69 & 0.51 & 0.33\\
		\hline
	\end{tabular}
	
	\label{planes}
\end{table}

Table \ref{planes} shows the model's highest training and testing results using input data sampled from different planes throughout the training process to predict $\sigma_{zz}$. All models showed overfitting to some degree, however, the model which was trained on data sampled from the xy-plane had the best performance with an $R^2$ score on the testing set of 0.69, indicating that data sampled from the xy-plane contained the most relevant information for predicting the corresponding values of $\sigma_{zz}$.

\begin{figure*}
    \centering
    \includegraphics[width=0.7\textwidth]{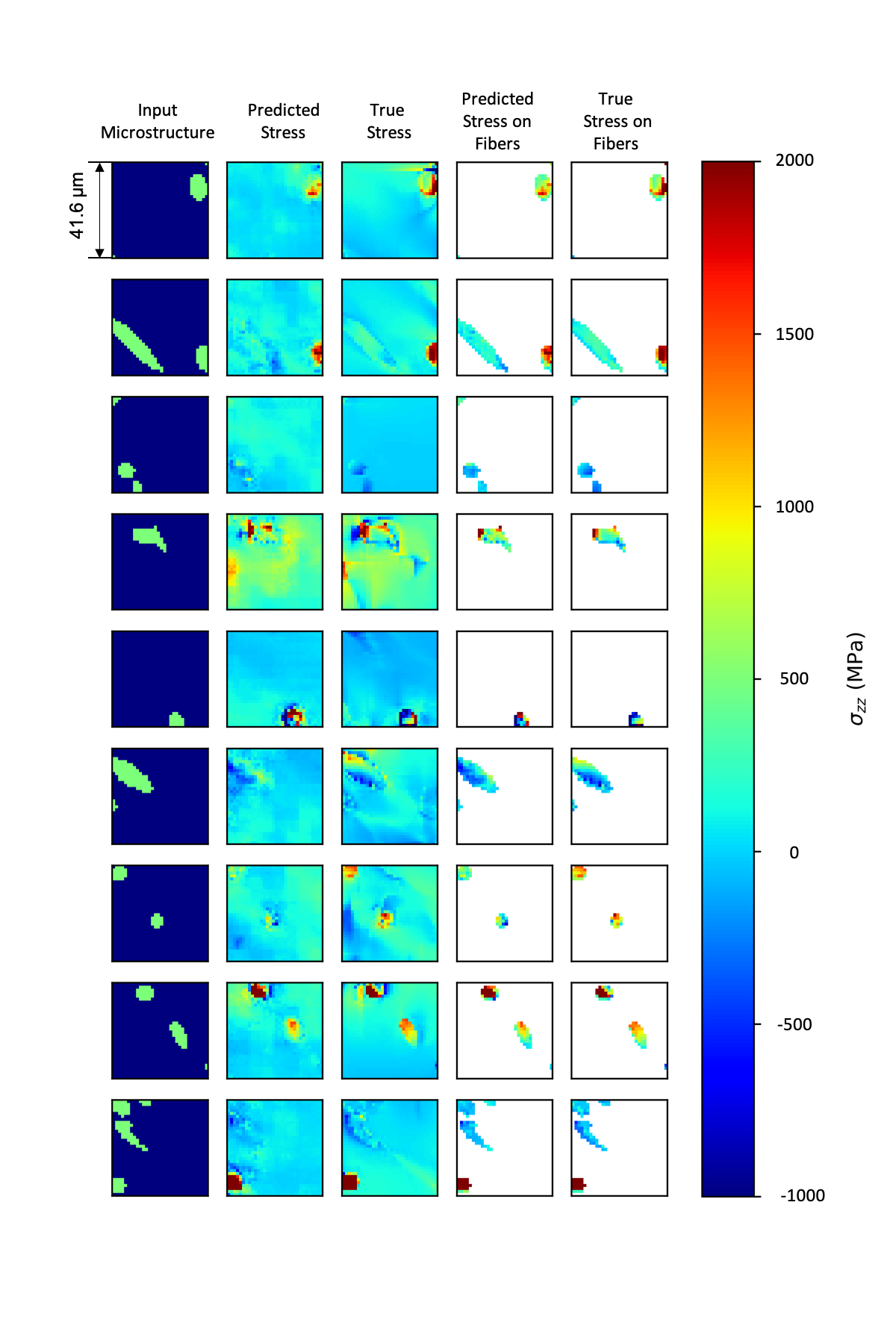}
    \caption{Visualization of the predicted stress fields from the CNN and the true stress fields from the FE simulation on the 9 data points randomly selected from the testing set. The first column shows the input microstructure; the second column is the corresponding predicted stress fields; the third column shows the true stress field obtained from FE simulation; the fourth and fifth columns show the predicted and true stress within the fibers, respectively. All stress metrics correspond to the normal stress relative to the loading direction, $\sigma_{zz}$.}
    \label{testing}
\end{figure*}

\begin{figure*}
    \centering
    \renewcommand*{\thesubfigure}{(\Alph{subfigure})}
    \subfigure[]{

		\includegraphics[width=\linewidth]{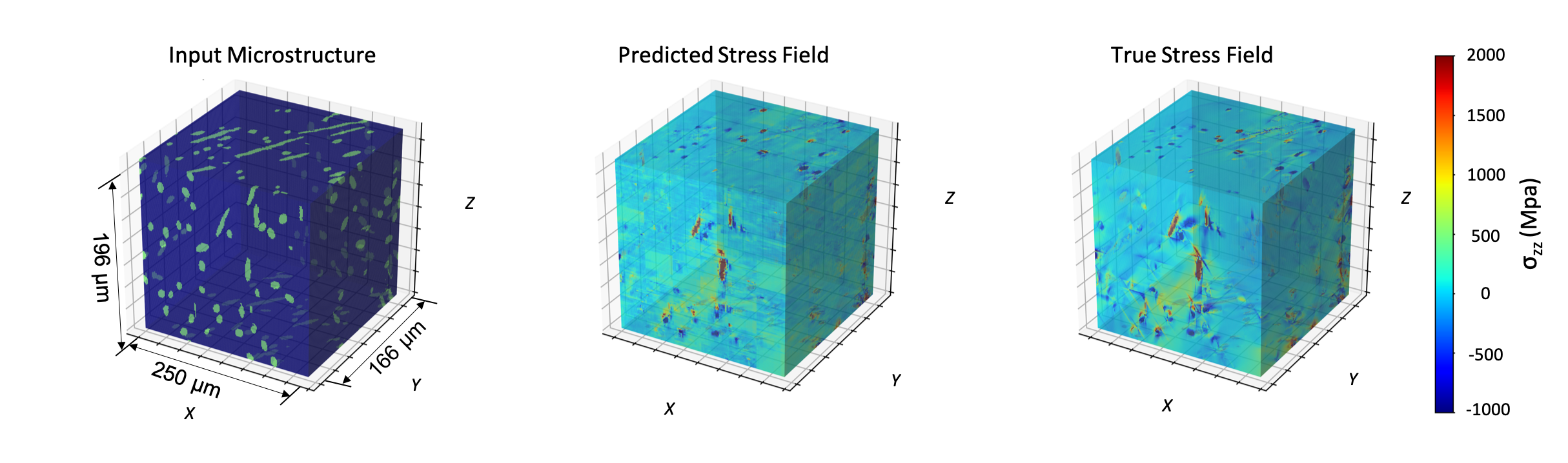}
		\label{fig:subfigure1}}
	\quad
	\subfigure[]{%
		\includegraphics[width=\linewidth]{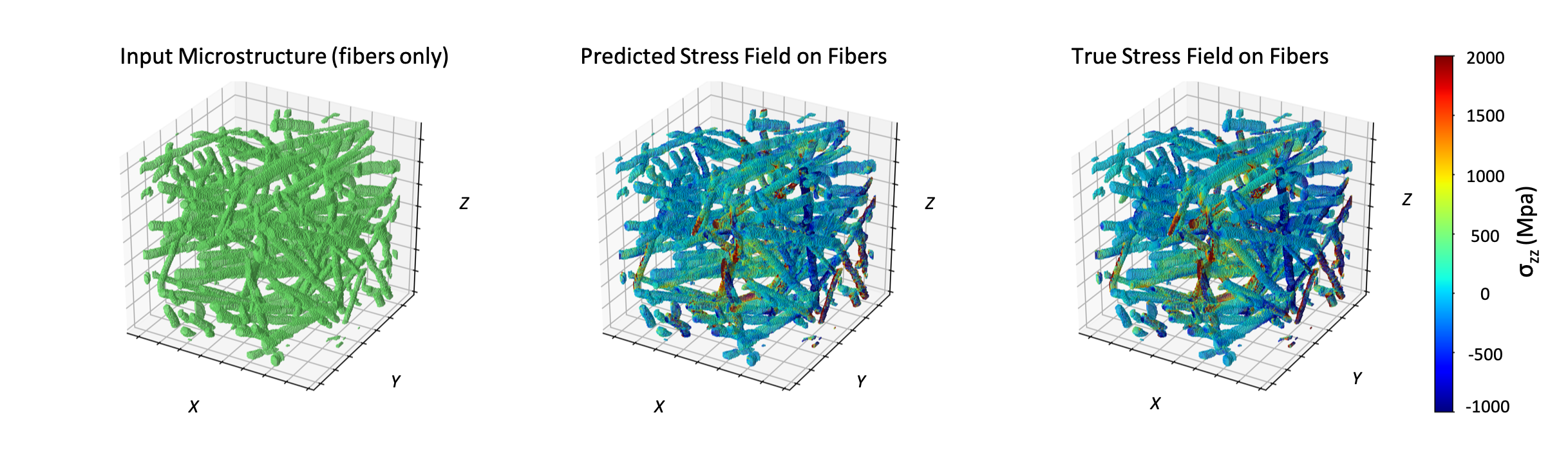}
		\label{fig:subfigure2}}
	\caption{(first column) 3D phase reconstruction of the microstructure. Reconstructed stress fields (normal stress along the loading axis) from (second column) the predicted by the CNN model and (third column) the FE simulation dataset representing the training and testing sets. (A) The stress field over the entire sample volume of the composite. (B) The stress field within the discontinuous glass fibers.}
	\label{reconstruction}
\end{figure*}

Although the testing $R^2$ of the best performing model (using input microstructural data sampled from the xy-plane) was not 1, it was able to capture the overall $\sigma_{zz}$ distribution. Figure \ref{testing} shows the visualization of the $\sigma_{zz}$ fields (predicted and true) on select slices in the xy-plane. The overall predicted $\sigma_{zz}$ on the testing data follows the corresponding stress fields, which were computed from the FE simulation, especially on areas with extreme values. The high-stress values consistently occurred within the fibers, and the trained model was able to accurately pinpoint the location where high stress was experienced within the fibers. Figure \ref{reconstruction} shows the reconstructed composite block and 3D fibers from the stacked slices used in the network training and testing. Compared to the stress fields obtained from the FE simulation, the network-learned stress fields had less sharp edges and lost some details. Nonetheless, the locations and magnitudes of extreme stress values were consistent between the network prediction and FE simulation, especially relative to the fibers. 




Some limitations exist in this study. Firstly, due to the high computational cost of the FE simulation, there was only one fiber-reinforced composite specimen available for network training and testing. This not only poses difficulties in training a network that has strong generalization ability, but it also creates challenges in validating the trained network on other composites. Secondly, no-load, displacement conditions or boundary conditions were taken into account as a part of the network input (as the CNN input was restricted to the stress fields results from the FE simulation, which did not include the model set-up or evolution of the stress fields during loading), which might have prevented the network from learning highly precise mapping to the stress field. It also limits the applicable scenarios of the trained network since it is only expected to work when predicting the stress field of a composite that is under the same conditions as the one used in training. Lastly, the slices were only sampled in 2-dimensions and from orthogonal planes, resulting in possible loss of information. These limitations are to be addressed in future studies.



\section{Conclusion}
\label{conclusion}

In this work, a fully convolutional neural network with an encoder-decoder structure was used to predict the stress field from the 2D microstructure slices of a fiber-reinforced polymer specimen. The CNN model was trained on FE simulations of an exact 3D microstructure, which demonstrated the areas of highest stress corresponded to damage initiation via an in-situ X-ray micro-computed tomography \cite{hanhan2020predicting}.  Hence, it is postulated as a corollary that the CNN model could identify regions of incipient damage relative to the microstructural features of a composite.  Further, the model was able to make fast predictions from the given 2D microstructure images via a single forward pass. The training and testing results showed that the network performed best on segmented microstructural images sampled from the xy-plane to predict the normal stress field in the z-direction, $\sigma_{zz}$. Although the $R^2$ statistic on the testing set was 0.69 (meaning 69\% of the variance in the true stress field could be explained by the trained network), the network was able to capture the important characteristics of the stress distribution, especially on fibers,  based on the visualized results. This network has proven to be useful in learning microstructure-stress mapping for non-linear fiber-reinforced polymers. It can help accelerate the evaluation of the structural integrity and potentially assist the identification of the locations where incipient damage formation could occur. Nevertheless, the trained network is only expected to work on the fiber-reinforced composites under the same settings as the specimen used in this study. To expand the generalization ability of a trained network, some future work can involve (1) adding load and various boundary conditions into the training process using the same network; (2) adopting Conditional Generative Adversarial Network (CGAN) \cite{mirza2014conditional} to learn stress fields conditioned on different boundary conditions so that new stress fields can be generated based on other boundary conditions; and (3) sampling 3D blocks and utilizing 3D convolution operation to extract features, which is expected to preserve more spatial information of the data points.

\section*{Acknowledgment}
The authors would like to thank Purdue’s Institute for Global Security and Defense Innovation for support for this work based on a Rapid Innovation Award.  Additionally, GL and YS gratefully acknowledge the support from the National Science Foundation (DMS-1555072, DMS-1736364, CMMI-1634832 and CMMI-1560834), and Brookhaven National Laboratory Subcontract 382247, ARO/MURI grant W911NF-15-1-0562 and U.S. Department of Energy (DOE) Office of Science Advanced Scientific Computing Research program DE-SC0021142. MDS and IH would like to thank the National Science Foundation CMMI MoM, Award No. 1662554 (Program Manager Dr. Siddiq Qidwai) for support.  Dr. Xianghui Xiao from Argonne National Laboratory and Ronald Agyei from Purdue University assisted in the micro-computed tomography characterization.  This research used resources of the Advanced Photon Source, a U.S. Department of Energy (DOE) Office of Science User Facility operated for the DOE Office of Science by Argonne National Laboratory under Contract No. DE-AC02-06CH11357. 

\section*{Declaration of Competing Interest}
The authors declare no competing financial interest.

\printcredits

\section*{Data Availability}

The processed data required to reproduce these findings are available to download from \url{https://github.com/sunyx1223/stress_fiber_polymer}.

\bibliographystyle{unsrt}

\bibliography{reference}

\end{document}